\documentclass{PoS}

\title{Lessons learnt from INTEGRAL AGN}

\ShortTitle{Lessons learnt from INTEGRAL AGN}

\author{\speaker{Volker Beckmann}$^{a}$, C. Ricci$^{b}$, S. Soldi$^{c}$, J. Alfonso-Garz\'on$^{d}$, T.J.-L. Courvoisier$^{b}$, 

A. Domingo$^{d}$, N. Gehrels$^e$, P. Lubi\'nski$^{f}$, J.M. Mas-Hesse$^{d}$, A.A. Zdziarski$^{f}$\\ 
	\llap{$^a$} Centre Fran\c{c}ois Arago, APC, IN2P3/CNRS, Universit\'e Paris Diderot, 10 rue Alice Domon et L\'eonie Duquet, 75205 Paris Cedex 13, France\\
        \llap{$^b$} ISDC Data Centre for Astrophysics, 16 Ch. d'Ecogia, 1290 Versoix, Switzerland
\\
	\llap{$^c$} Laboratoire AIM - CNRS - CEA/DSM - Universit\'e Paris Diderot (UMR 7158), CEA Saclay, DSM/IRFU/SAp, 91191 Gif-sur-Yvette, France\\
        \llap{$^d$} Centro de Astrobiologia LAEX (CSIC-INTA), 28691 Villanueva de la Canada, Madrid \\
        \llap{$^e$} Astrophysics Science Division, Code 661, NASA Goddard Space
        Flight Center, MD 20771, USA\\
        \llap{$^f$} Centrum Astronomiczne im. M. Kopernika, PL-00-716 Warszawa, 
Poland\\
        E-mail: \email{beckmann@apc.univ-paris7.fr}}
 
\abstract{The {\it INTEGRAL} mission provides a large data set for studying the hard X-ray properties of AGN and allows to test the unified scheme for AGN. We present results based on the analysis of 199 AGN. A difference between the Seyfert types is detected in slightly flatter spectra with higher cut-off energies and lower luminosities for the more absorbed/type 2 AGN. When applying a Compton reflection model, the underlying continua ($\Gamma \simeq 1.95$) appear the same in Seyfert 1 and 2, and the reflection strength is $R \simeq 1$ in both cases, with differences in the inclination angle only. A difference is seen in the sense that Seyfert~1 are on average twice as luminous in hard X-rays than the Seyfert~2 galaxies. 
The unified model for Seyfert galaxies seems to hold, showing in hard X-rays that the central engine is the same in Seyfert 1 and 2 galaxies, seen under different inclination angle and absorption. 
Based on our knowledge of AGN from {\it INTEGRAL} data, we briefly outline open questions and investigations to answer them. In this context an ultra-deep ($\gae 12 \rm \, Ms$) extragalactic field can be a true legacy of the {\it INTEGRAL} mission in the area of AGN studies.
}

\FullConference{The Extreme sky: Sampling the Universe above 10 keV\\
                 October 13-17 2009\\
                 Otranto (Lecce) Italy}

\newcommand{\gae}{\mathrel{>\kern-1.0em\lower0.9ex\hbox{$\sim$}}}
\begin{document}

\section{Introduction}
A longstanding discussion has been, whether Seyfert 1 and Seyfert 2 galaxies (or un-absorbed and absorbed Seyferts) indeed
represent physically different types of objects, or if they can
be unified under the assumption that they are intrinsically the same
but seen from a different viewing angle with respect to absorbing
material in the vicinity of the central engine, 
and that the difference in X-ray spectral slope can be explained
  solely by the absorption and reflection components. This {\it unified model} naturally explains the different
Seyfert types in a way that the broad-line region is either visible
(Seyfert~1) or hidden (Seyfert~2) possibly by the same material in
the line of sight that is responsible for the absorption detectable at soft
X-rays. 
On the other hand, the model has some problems explaining other aspects of AGN, for example, that some Seyfert galaxies change their type from 1 to 2 and back \cite{Cohen86,Aretxaga99}. 
Also, the existence of
Seyfert~2 galaxies that show no absorption in the soft X-rays, like
NGC~3147 and NGC~4698 \cite{Pappa01} cannot be explained by the unified model.
The AGN surveys provided by {\it Swift}/BAT \cite{Tueller08} and
{\it INTEGRAL} IBIS/ISGRI
\cite{XLF,Bassani07} have already led to the
discovery that the fraction of absorbed and Compton thick sources is
less than expected from cosmic X-ray background synthesis models
(e.g. \cite{Treister05}). With the ongoing
{\it INTEGRAL} mission, it is now possible to compile a large sample of AGN
for spectroscopic and correlation studies \cite{Beckmann09} and to probe the
 unified model for AGN.

\section{The intrinsic hard X-ray spectrum}

The effect that Seyfert 1 and low-absorbed objects appear to have steeper
X-ray spectra than the Seyfert 2 and highly absorbed AGN was first
noticed based on {\it Ginga} and {\it CGRO}/OSSE data \cite{Sy1average} and later
 confirmed by various other studies.
The difference in the hard X-ray spectral slope between Seyfert 1 and 2
 has been a point of discussion ever since its discovery. 

A solution might be provided when considering the effects of Compton
reflection on the hard X-ray spectrum. Recent analysis of a sample of 105 Seyfert galaxies using the spectra
collected
with {\it BeppoSAX} in the 2--200 keV band \cite{Dadina08} provided no evidence
of any spectral slope difference when applying more complex model
fitting including a reflection component (PEXRAV). The mean photon index values
found for Seyfert 1 and Seyfert 2 samples were $\Gamma = 1.89 \pm 0.03$ and
$\Gamma = 1.80 \pm 0.05$. The difference between types 1 and 2 is seen in
this model in the different strength of the reflection component, with
$R = 1.2 \pm 0.1$ and $R = 0.9 \pm 0.1$, and different cut-off
energies of $E_C = 230 \pm 22 \rm \, keV$ and $E_C = 376 \pm 42 \rm \,
keV$, for Seyfert~1 and Seyfert~2, respectively. It has to be
pointed out that spectral slope, reflection strength, and cut-off
energy are closely linked.   
The IBIS/ISGRI data have a
  disadvantage over broad-band data when studying 
  the spectral shape of the hard X-ray continuum, as they
  lack information about the spectrum below 18 keV. But on the
  positive side there is no influence by data taken at different epochs. Even when e.g. {\it Swift}/XRT and {\it INTEGRAL} data are taken simultaneous, the XRT data over only a very small fraction of the total IBIS/ISGRI exposure time, usually of the order of a few per cent. 
  In those cases, where spectra are taken
  by more than one instrument at different times, the flux variability can
  mimic a stronger or weaker reflection component or cut-off energy. Although the spectra of AGN as seen by IBIS/ISGRI above 20 keV show little variability, significant changes in flux and spectral shape are indeed seen in the energy range below 10 keV. 

The {\it INTEGRAL} data show consistent slopes for the spectra
of unabsorbed / type 1 and absorbed / type 2 objects already
when a simple power-law model is used \cite{Beckmann09}.
When applying the model used by Dadina (2007) to the stacked
 {\it INTEGRAL} spectrum of Seyfert galaxies, we get similar results:
 the underlying powerlaw appears to have consistent (within $2\sigma$) spectral slope
 for type 1 ($\Gamma = 1.96$) and type 2 ($\Gamma = 1.91$) objects
 and the same reflection strength $R \simeq 1.1$, when applying different inclination angles of $i \simeq 30^\circ$ and
 $i \simeq 60^\circ$, respectively.
The data do not allow to determine the cut-off energy or
inclination angle when fitting the reflection component. When fitting a simple cut-off power law, the
{\it INTEGRAL} data show the same trend as the {\it BeppoSAX} sample,
i.e. a lower cut-off energy for Seyfert 1 ($E_C = 86 \rm \, keV$) than
for Seyfert 2 ($E_C = 184 \rm \, keV$). It has to be taken
  into account, though, that the fit to the Seyfert~2 data is of bad
  quality, and that fixing the cut-off here to the same value as derived for the Seyfert~1, also leads to the same spectral slope. 
\begin{table}
\caption[]{Results from spectral fitting of a simple power law, a cut-off power law, and a Compton reflection model (PEXRAV) to the stacked IBIS/ISGRI
 spectra of {\it INTEGRAL} AGN}
\begin{tabular}{lcccc}
\noalign{\smallskip}
\hline\hline
sample    & $\Gamma$ & $E_C \, [\rm keV]$ & $R$ & $\chi^2_\nu$\\
\hline
Sey~1 ($\ge 5\sigma$) & $1.96 {+0.03 \atop -0.02}$ & -- & -- & 5.66\\
                   & $1.44 \pm 0.10$ & $86 {+21 \atop -14}$ & -- &
                   1.10\\
\, \, \, \,$(i = 30^\circ)$   & $1.96 \pm 0.02$ & -- & $1.2 {+0.6 \atop -0.3}$ &
1.15\\
\hline
Sey~1.5 ($\ge 5\sigma$) & $2.02 \pm 0.04$ & -- & -- & 3.54\\
                   & $1.36 \pm 0.15$ & $63 {+20  \atop -12}$ & -- &
                   0.57\\
\,\, \, \,  $(i = 45^\circ)$   & $2.04 \pm 0.04$ & -- & $3.1 {+4.7 \atop -1.3}$
                   & 0.29\\
\hline
Sey~2 ($\ge 5\sigma$) & $1.89 {+0.04 \atop -0.02}$ & -- & -- & 3.13\\
                   & $1.65 \pm 0.05$ & $184 {+16 \atop -52}$ & -- &
                   2.58\\
\,\,  \, \, $(i = 60^\circ)$   & $1.91 {+0.02 \atop -0.03}$ & -- & $1.1  {+0.7 
                   \atop -0.4}$ & 1.67\\
\hline
all Sey ($\ge 5\sigma$) & $1.97 \pm 0.02$ & -- & -- & 6.18\\
                   & $1.44 {+0.08 \atop -0.13}$ & $86 {+16 \atop
                     -17}$ & -- & 1.96\\
\,\, \, \,  $(i = 45^\circ)$   & $1.95 \pm 0.02$ & -- & $1.3 {+0.7 \atop
                     -0.4}$ & 1.53\\
\hline
Unabs. ($\ge 5\sigma$) & $1.97 {+0.03 \atop -0.01}$ & -- & -- & 5.64\\
                   & $1.53 {+0.09 \atop -0.08}$ & $100 {+25 \atop -15}$ & -- &                    1.82\\
\,\,  \, \,  $(i = 45^\circ)$    & $1.98 \pm 0.02$ & -- & $1.3 {+0.6 \atop -0.4}$ & 0.95\\ 
\hline
Abs. ($\ge 5\sigma$) & $1.91 {+0.04 \atop -0.03}$ & -- & -- & 1.2\\
                   & $1.43 {+0.13 \atop -0.08}$ & $ 94 {+32 \atop -13}$ & -- &                    1.59\\
\,\,  \, \, $(i = 45^\circ)$    &  $1.91 {+0.02 \atop -0.03}$ & -- & $1.5 {+1.5 \atop -1.4}$ & 1.03\\
\end{tabular}
\label{spectralfits}
\end{table}
When fitting a reflection model to the stacked data, one gets a consistent
photon index of $\Gamma \simeq 1.95$ and reflection strength $R
\simeq 1.3$
for both absorbed and unabsorbed AGN. Table~\ref{spectralfits} summarises the results of the spectral fitting to the stacked ISGRI spectra of the various Seyfert subtypes.
The values of $\Gamma$ and $R$ from the PEXRAV spectral modeling agree with the
correlation $R = (4.54 \pm 1.15) \times \Gamma - (7.41 \pm 4.51)$
found for the {\it BeppoSAX} AGN sample \cite{Dadina08}, which, for the
{\it INTEGRAL} sample with $\Gamma = 1.95$, would lead to $R = 1.4$. This
$R(\Gamma)$ correlation was first noted based on {\it Ginga} data for
extragalactic and Galactic black holes,
leading to $R = (1.4 \pm 1.2) \times 10^{-4} \, \Gamma^{(12.4 \pm 1.2)}$
\cite{Zdziarski99}, which in our case would result in a smaller
expected reflection component with $R = 0.6$ but within $1 \sigma$ of
the value detected here. Absorbed and unabsorbed sources show a consistent
turnover at about $E_C = 100 \rm \, keV$ when a cut-off power law
model is applied.

The observed dichotomy of different spectral slopes for
type 1 and type 2 objects might therefore be caused by data with too
low significance, which do not allow to fit the reflection component, 
or in general by a strong dependence of the spectral slope on the choice of the
fitted model.
One aspect that has to be kept in mind is the dependence
of the reflection strength $R$ on the model applied and on the
geometry assumed. 
Murphy \& Yaqoob (2009) showed recently that their model of a reflection spectrum from
a Compton-thick face-on torus that subtends a solid angle of $2\pi$
at the X-ray source is a factor of $\sim 6$ weaker than that expected
from a Compton-thick disc as modelled in PEXRAV. Therefore,
applying this torus model to the data presented here would result in a much
larger relative reflection
strength. 

A difference between type 1 and type 2 objects is seen in the
average luminosity of these subclasses. For 60 absorbed Seyfert
galaxies, the average luminosity is $\langle L_{20 - 100 \rm keV} \rangle = 2.5
\times 10^{43} \rm \, erg \, s^{-1}$, more than a factor of 2
lower than for the 74 unabsorbed Seyfert with redshift information ($\langle L_{20 - 100
 \rm keV}\rangle  = 6.3 
\times 10^{43} \rm \, erg \, s^{-1}$). 
The differences in luminosities are exactly the
 same when excluding the 5 Compton thick objects.

\section{Open questions}

{\it INTEGRAL} data have already led to a number of results concerning AGN at hardest X-rays. The local luminosity function seems to be well established now, the source population at this energy and spectral characteristics are now well described. Further observations, which might lead to a doubling of the number of {\it INTEGRAL} AGN will not change significantly the results in this field. 

There are several investigations that can be done exclusively by {\it INTEGRAL}.
\begin{itemize} 
\item as it has been shown in Tab.~\ref{spectralfits}, the cut-off energy and reflection strength cannot be fit simultaneously when using the stacked spectrum of all {\it INTEGRAL} AGN. It also shows that the cut-offs in AGN spectra appear to be mainly at high energies, i.e. >80 keV. In order to constrain the cut-off, IBIS/ISGRI data on the brightest Seyferts are needed. This can be done using the brightest sources of the {\it INTEGRAL} AGN catalogue \cite{Beckmann09} as it has been shown already in several investigations (e.g. \cite{Beckmann08,Molina09}).
The {\it INTEGRAL} community should aim at completing pointed observations on the brightest AGN in order to form a complete sample.
\item to measure the evolution of the {\it INTEGRAL} detectable AGN it is necessary to perform an ultra-deep ($\gae 12 \rm \, Ms$) extragalactic field observation. Several medium-deep surveys have already been performed, e.g. in the XMM-LSS field \cite{Virani10}, in the Virgo region \cite{Paltani08}, and on the north ecliptic pole. {\it INTEGRAL} observation time should now be efficiently used in order to extend one of these into a ultra-deep field, which will be able to determine the evolution in redshift. Expectations for such a field range from 40 to 110 detectable AGN, depending on the assumptions and the evolutionary behaviour. This homogeneous sample would also be suitable to settle the debate about the true fraction of Compton thick sources among the hard X-ray detected AGN.
\item detailed variability studies can be performed using a small well-defined sample of the brightest AGN, which have sufficient long exposures and/or several observation periods and combining the {\it INTEGRAL} data with those collected by other missions. Few studies like this have been performed, e.g. on 3C 273 \cite{Soldi08}, and further work is in progress \cite{Soldi10}. 
\item the question of spectral variability at hardest X-rays can also be addressed uniquely by {\it INTEGRAL} using the brightest detected AGNs. Such studies have been performed e.g. on 3C 273 \cite{Masha07}, NGC~4151 \cite{Lubinski10}. Other bright objects, e.g. NGC 4945 and IC 4329A, are still to be investigated. In this context, short (200 ksec) follow-up observations on the $\sim 10$ brightest variable AGN would be useful in order to study their long term (on years time scale) spectral behaviour.
\item {\it INTEGRAL} observations of blazars in outburst provide the highest quality hard X-ray spectra and help to constrain the spectral energy distribution \cite{3C454Pian,BeckmannFermi}. This is especially important as {\it Fermi}, {\it AGILE}, and TeV experiments rely heavily on simultaneous multiwavelength observations. 
\end{itemize}

\section{Conclusions}

 The AGN population detected by {\it INTEGRAL} is
dominated by Seyfert galaxies in the local ($\langle z \rangle =
0.03$) universe, with moderate X-ray luminosity ($\langle L_{20 - 100
 \rm keV} \rangle = 4 \times 10^{43} \rm \, erg \, s^{-1}$). Taking into account the results on the Eddington ratios of the {\it INTEGRAL} detected AGN \cite{Beckmann09}, Seyfert~1 galaxies appear to
have 
higher luminosities ($\langle L_{20 - 100
 \rm keV} \rangle = 10^{44} \rm \, erg \, s^{-1}$) and Eddington ratio
($\langle \lambda_{\rm Sy1} \rangle = 0.064$) than the Seyfert~2 galaxies 
($\langle L_{20 - 100 \rm keV} \rangle = 2.5 \times 10^{43} \rm \, erg \, s^{-1}$,
$\langle \lambda_{\rm Sy2} \rangle = 0.02$). 
The underlying continuum of the hard X-ray spectrum appears to be consistent
between different Seyfert types, both when a simple power-law model
is applied and 
 when considering the effects of Compton
 reflection. 
Applying the PEXRAV reflection model with no high-energy
 cut-off, the Seyfert~1 and 2 galaxies show the same underlying power
 law with $\Gamma \simeq 1.95$ and a reflection component of $R \simeq
 1.1$. 
On the other hand, when applying a cut-off power law model to the stacked spectra, the Seyfert~1 show lower
cut-off energies ($E_C = 86 {+21 \atop -14} \rm \, keV$) than the
Seyfert~2 objects ($E_C = 184 {+16 \atop -52} \rm \, keV$). However, the bad
quality of the fit in the latter case and the fact that fixing the
cut-off to the value of the Seyfert~1 leads to a similar spectral
slope might indicate that the spectra are intrinsically indeed the same.
The same differences 
as for different Seyfert classes 
are observable when considering the intrinsic absorption: the unabsorbed sources also have 
higher luminosities ($\langle L_{20 - 100 \rm keV}\rangle  = 6.3
\times 10^{43} \rm \, erg \, s^{-1}$) and Eddington ratio
($\langle \lambda \rangle = 0.06$) than the absorbed AGN 
($\langle L_{20 - 100 \rm keV} \rangle = 2.5
\times 10^{43} \rm \, erg \, s^{-1}$, $\langle \lambda \rangle =
0.015$).

The overall picture can be interpreted within the scenario of a
unified model.
The whole hard X-ray detected Seyfert population fills the parameter
space of spectral shape, luminosity, and accretion rate smoothly, and
only an overall tendency is seen in which more massive objects are
more luminous, less absorbed, and accreting at higher Eddington ratio.
More evidence for the unified scheme is that a fundamental plane can be found between the mass of the central
object and optical and X-ray luminosity \cite{Beckmann09}. The correlation takes
  the form $L_V \propto L_X^{0.6} M_{BH}^{0.2}$,
 similar to what is found in previous studies between radio luminosity $L_R$, $L_X$, and $M_{BH}$ \cite{Merloni03}. This links the
accretion mechanism with the bulge of the host galaxy and with the
mass of the central engine in the same way in all types of Seyfert
galaxies. 
Further exploitation of the data is necessary to study the spectral characteristics of a bright sub-sample in a consistent way. In addition, variable and bright Seyferts can be used to study spectral variability by performing additional short ($\sim200 \rm \, ks$) observation with IBIS/ISGRI in the upcoming AOs.  
Evolutionary effects cannot be studied by {\it INTEGRAL} at the time being. Only if an ultra-deep ($\gae 12$ Ms) extragalactic field study is performed, we will be able to clarify how the hard X-ray AGN fraction evolves with redshift. Such a study would represent one of the important legacies of the {\it INTEGRAL} mission.

\end{document}